\documentclass[a4paper,pre,twocolumn,amsmath,amssymb,longbibliography]{revtex4-1}
\usepackage{bm}
\usepackage{graphicx,color}
\usepackage{color,soul}
\usepackage[colorlinks,linkcolor=magenta,citecolor=magenta]{hyperref}

\newcommand{\mean}[1]{\langle #1 \rangle}

\newcommand{\lam}{\lambda}

\newcommand{\kT}{k_\text{B}T}

\AtBeginDocument{%
    \newwrite\bibnotes
    \def\bibnotesext{Notes.bib}
    \immediate\openout\bibnotes=\jobname\bibnotesext
    \immediate\write\bibnotes{@CONTROL{REVTEX41Control}}
    \immediate\write\bibnotes{@CONTROL{%
    apsrev41Control,author="08",editor="1",pages="1",title="1",year="1"}}
    \if@filesw
    \immediate\write\@auxout{\string\citation{apsrev41Control}}%
    \fi
}%

\begin{document}

\title{Coarse-grained models for phase separation in DNA-based fluids}

\author{Soumen De Karmakar}
\author{Thomas Speck}
\affiliation{Institute for Theoretical Physics IV, University of Stuttgart, Heisenbergstr. 3, 70569 Stuttgart, Germany}


\begin{abstract}
  DNA is now firmly established as a versatile and robust platform for achieving synthetic nanostructures. While the folding of single molecules into complex structures is routinely achieved through engineering basepair sequences, much less is known about the emergence of structure on larger scales in DNA fluids. The fact that polymeric DNA fluids can undergo phase separation into dense fluid and dilute gas opens avenues to design hierachical and multifarious assemblies. Here we investigate to which extent the phase behavior of single-stranded DNA fluids is captured by a minimal model of semiflexible charged homopolymers while neglecting specific hybridization interactions. We first characterize the single-polymer behavior and then perform direct coexistence simulations to test the model against experimental data. We conclude that counterions not only determine the effective range of direct electrostatic interactions but also the effective attractions.
\end{abstract}

\maketitle


\section{Introduction}

Deoxyribonucleic acid (DNA) is traditionally known for carrying the genetic information in living organisms. It also is an excellent \emph{material} and has been extensively used for the fabrication of synthetic nanostructures \cite{jones15, huYong19, zhan23}, harnessing the ability of nucleobases to form highly specific and stable bonds. Three principle strategies have emerged: First come tile-based DNA assemblies \cite{seeman03}. Second, DNA origami \cite{shih04, rothemund06, hong17, dey21} employs long, single-stranded DNA (ssDNA) molecules known as \emph{scaffold} together with specifically designed short \emph{staple} oligonucleotides to realize a plethora of two- and three-dimensional nanostructures through \emph{folding} that have found numerous applications from drug delivery devices \cite{hu19} and plasmonic engines \cite{liu18} to DNA memory devices \cite{doricchi22}. And third, DNA molecules coupled with inorganic and organic nanoparticles \cite{mirkin96, lowensohn20} and polymers \cite{schnitzler12, luckerath20, li21} is another avenue to realize various nanostructures. Moreover, similar strategies can been followed replacing DNA by RNA \cite{grabow14, li15}. Despite the tremendous success of DNA to create static and dynamic nanostructures, its scalability towards larger hierarchical mesostructures has proven challenging \cite{merindol18}. One particularly exciting application is the autonomous formation of synthetic cell-mimetic structures that can be endowed with biologically relevant function \cite{samanta20, samanta22, malouf23, leathers22, kohyama22}.

A promising direction is the initial formation of larger aggregates from disordered fluid DNA molecules that subsequently adopt nanoscale order programmed into their nucleobase sequences. Such phase separation of macromolecules is well understood for synthetic block copolymers \cite{bates91, deSilva17, kohno15, qiu15, groschel12}. In living cells, it is now widely accepted that phase separation underlies the formation of certain membraneless organelles through biomolecular condensates \cite{julicher24}, typically associated with intrinsically disordered proteins \cite{bauer22}. Frequently, the amphiphilic nature of synthetic polymers \cite{zumbro21} and low complexity proteins \cite{devarajan24, dignon18, dignon19, mabuchi23} provides the effective interactions necessary to initiate phase separation. The hydrophobic nature of the nucleic acid backbone has been recently unveiled \cite{merindol18, wadsworth23}; exhibiting phase separation beyond a lower critical solution temperature (LCST) \cite{dignon19}. Combining such LCST-type phase separation in DNA with the hybridization of DNA nucleobases has already shown promising results for building hierarchical protocell-like structures at the micro scale and beyond that were inaccessible before \cite{merindol18, samanta22, liu22}.

Although temperature-sensitive hydrophobic interactions between the nucleic acid backbones is believed to be the primary driving mechanism for the observed LCST phenomenon (through partial release of their hydration shells), the roles of different nucleobases have not been explored systematically yet. For instance, ssDNA comprising a sufficiently large block of adenine has been reported to exhibit LCST-type phase transition in which aggregates resolve quickly after cooling. Replacing adenine with guanine also leads to a transition but now aggregates remain stable and do not dissolve even when the external temperature is decreased well below the cloud-point temperature \cite{merindol18}. Moreover, apart from shielding the electrostatic charges of the nucleic acid backbone, multivalent background counterions might play an important role by bridging the intra- and inter-molecular nucleotides, as recently demonstrated in the work of Wadsworth \emph{et al.} \cite{wadsworth23}. A comprehensive understanding of the underlying molecular interactions is not only advantageous from a materials point of view but eventually might lead to cures for certain neurological diseases that are related to changes in the phase behavior of various biomolecular condensates, e.g., solidification, with time \cite{patel15}.

Here we investigate a minimal coarse-grained model of ssDNA (and ssRNA) homopolymers (i.e., only one type of nucleobase) as a first step towards a computational platform that exploits LCST-type phase separation in ssDNA fluids for designing mesoscale structures. Unlike most of the existing high \cite{guerra24, ouldridge11, snodin15} and low \cite{largo07, lara11, knorowski11, ghobadi16, prhashanna19} resolution DNA models that are used to investigate structural details, our model is intended to investigate large-scale fluid-like phase behavior. We thus opt for a description in which two nucleotides are combined into one bead, comparable to residue-resolved models such as the hydrophobicity scale (HPS) model for intrinsically disordered proteins~\cite{dignon18}.

Our minimal model (Sec.~\ref{sec:model}) is guided by models for semiflexible charged polymers \cite{fazli09, sayar10, hao18}. The model employs generic pair potentials for bonded and non-bonded interactions, for which we follow a \emph{top-down} strategy to determine the parameters based on geometric considerations and experimental data. To this end, in Sec.~\ref{sec:params} we first investigate single-molecule features over a wide range of counterion concentrations and valencies. We then incorporate an additional attractive non-bonded pair potential and investigate the coil-to-globule transition of a single homopolymer in Sec.~\ref{sec:coilGlobTran}. Finally, we turn to direct coexistence simulations of many polymers (Sec.~\ref{sec:phaseTran}) and map the resulting phase diagram of our model fluid to available (still scarce) experimental data \cite{merindol18, wadsworth23}.


\section{Model}
\label{sec:model}

\begin{figure}[b!]
  \includegraphics{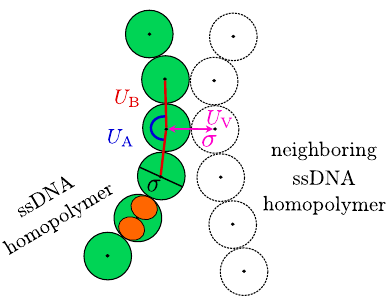}
  \caption{\label{fig:sketch} Diagram of the ssDNA homopolymers. Coarse-grained beads (green) of diameter $\sigma$ are connected by the harmonic potential $U_\text{B}$ to form the single strands. Each bead represents two nucleotides (orange). Flexibility is controlled through the three-body angle potential $U_\text{A}$ between three consecutive beads along the ssDNA. A neighboring ssDNA molecule is shown on the right (dashed beads). Excluded volume interaction between the intra- or intermolecular beads is modeled by the WCA potential $U_\text{V}$.} 
\end{figure}

We employ a model in which each ssDNA homopolymer is composed of $N$ coarse-grained beads with diameter $\sigma$ and mass $m$, where each bead represents two nucleotides (cf. Fig.~\ref{fig:sketch}). Neighboring beads are connected by a harmonic bond potential $U_\text{B}(r_{ij}) = K_\text{B} (r_{ij} - r_0)^2$ with $r_{ij}$ the distance between two beads, centered at $\boldsymbol{r}_i$ and $\boldsymbol{r}_j$, and $r_0$ is the bond rest length. The bond strength is fixed at a large value $K_\text{B} = 8530$ (50 kcal mol$^{-1}$ \AA$^{-2}$) to ensure rigid bonds \cite{kapoor24}. For the simulations, we report dimensionless values in units of the length scale $\sigma$, energy scale $\kT$ with $T=298$ K, and time scale $\tau=\sqrt{m \sigma^2/\kT}$ with bead mass $m$. Explicit units are provided in appropriate places when referring to experiments.

The excluded volume of beads is modeled through a repulsive Weeks-Chandler-Andersen (WCA) \cite{weeks71} pair potential
\begin{equation}
  U_\text{V}(r_{ij}) = \begin{cases}
    U_\text{LJ}(r_{ij};\epsilon) + \epsilon & r_{ij} \leq 2^{1/6}\sigma \\
    0 & r_{ij} > 2^{1/6}\sigma
  \end{cases}
  \label{eq:wca}
\end{equation}
based on the conventional Lennard-Jones potential
\begin{equation}
  U_\text{LJ}(r;\epsilon) = 4 \epsilon \left[\left(\frac{\sigma}{r}\right)^{12} - \left(\frac{\sigma}{r}\right)^6\right].
  \label{eq:lj}
\end{equation}
We will use $\sigma$ as the basic length scale. Geometric considerations of two neighboring molecules implies $\sigma\simeq1$ nm (averaging over stacked and unstacked configurations) \cite{seeman03}. Throughout, we fix the strength $\epsilon$ of the volume exclusion at $\epsilon = 6.8$ \cite{kapoor24}. Since ssDNA is a semiflexible polymer, we constrain bending through the three-body angle potential $U_\text{A} (\theta_{ijk})= K_\text{A} (\cos \theta_{ijk} - \cos \theta_0)^2$ between three consecutive beads, where $\theta_{ijk} = \theta_{ij} - \theta_{jk}$ is the bond angle and $\theta_0 = \pi$. The strength $K_\text{A}$ is a free parameter in our model and we will investigate its role below.

In addition to volume exclusion and bonded interactions, we consider two non-bonded interactions. First, electrostatic interactions are accounted for through the pair potential
\begin{equation}
  U_\text{C}(r_{ij}) = \kT\lam_\text{B}q_iq_j\frac{e^{-\kappa r_{ij}}}{r_{ij}}
\end{equation}
with partial charges $q_i$ carried by bead $i$. The solvent sets the Bjerrum length $\lam_\text{B}=e^2/(4\pi\epsilon_0\epsilon_\text{r}\kT)$, where $e$ denotes the magnitude of the electronic charge, $\epsilon_0$ is the permittivity of free space, and $\epsilon_\text{r}$ is the dielectric constant of the solvent. For water at room temperature, one finds $\lam_\text{B}\simeq0.71$~nm. The simplest theoretical approach, mean-field Debye-Hückel theory, predicts that charges in solution are screened with inverse screening length
\begin{equation}
  \kappa = \sqrt{8\pi\lam_\text{B} N_\text{A}I}
  \label{eq:kappa}
\end{equation}
due to freely mobile counterions, where $I$ is their ionic strength measured in mM (millimolar) and $N_\text{A}$ is Avogadro's number. Second, an effective attractive interaction $U_\text{H}(r)=U_\text{LJ}(r;\epsilon_\text{H})$ between beads is included through the Lennard-Jones potential [Eq.~\eqref{eq:lj}] with binding free energy $\epsilon_\text{H}$. This term subsumes all other interactions, in particular due to hydrophobic and ion-mediated effects.

In this work, we employ stochastic Langevin dynamics simulations with friction coefficient $\gamma = 1$ using the LAMMPS molecular dynamics package (June 2022 version) \cite{thompson22, intveld08, shire21}. All simulations are performed at fixed temperature (corresponding to room temperature $T=298$ K). Considering $N = 125$ beads (250 nucleotides) of mass $m = 2 m_0$, where $m_0$ is the average mass of the DNA nucleotides ($m_0 \approx 307$ u), $\gamma = 1$ corresponds to a diffusion coefficient $D \approx 5 \times 10^{-6}$ cm$^{2}$ s$^{-1}$ at room temperature, which is estimated using the Stokes-Einstein-Sutherland relation $D = \gamma\kT/m$ \cite{baer24, tinland97}. To put this value into context, a previous experiment measured for ssDNA with 280 nucleotides a diffusion coefficient $D \approx 1.9 \times 10^{-7}$ cm$^{2}$ s$^{-1}$ that is about 25 times smaller \cite{tinland97}. Since we are not interested in dynamic properties, and since a small value for the friction coefficient was found to improve the sampling efficiency in the simulations, we will proceed with $\gamma = 1$.


\section{Results and discussion}

\subsection{Persistence length}
\label{sec:params}

\begin{figure}[b!]
  \centering
  \includegraphics{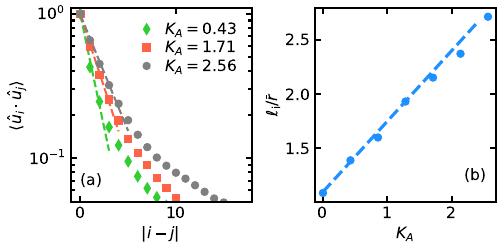}
  \caption{\label{fig:paraOpti1} Flexibility of single uncharged polymers ($q = 0$ and $\epsilon_\text{H} = 0$). (a)~Bond correlations $C(|i-j|) = \langle \hat{u}_i \cdot \hat{u}_j \rangle$ [Eq.~\eqref{eq:persisLen}] of two unit bond vectors separated by $|i - j|$ bonds for three different values of $K_\text{A}$ (see legend) and $r_0 = 1$. The dashed lines are exponential fits to the simulation data in the limit of small bond separation $|i - j|$. (b)~Intrinsic persistence length $\ell_\text{i}$ as a function of the strength of the angle potential $K_A$ at fixed $r_0 = 1$ (symbols). The dashed line is a linear fit.}
\end{figure}

The model parameters are: $\sigma$, $\epsilon$, $\epsilon_\text{H}$, $K_\text{B}$, $r_0$, $K_\text{A}$, $\kappa$, and the partial charges $q$. We have already fixed the values for $\epsilon$ and $K_\text{B}$, which we expect to play only a minor role for the collective behavior. For now we set $\epsilon_\text{H}$ to zero and study a single ssDNA chain. We consider a single ssDNA homopolymer of $N = 125$ beads in a cubic simulation box of length $L = 50$ with periodic boundaries. We randomly initialize a single molecule inside the simulation box at $t = 0$ and equilibrate the system for $t = 10000$ before collecting data for another $t = 10000$ at a uniform interval $\Delta t = 10$.

The persistence length $\ell=\ell_\text{i}+\ell_\text{c}$ is a measure for the flexibility of single chains. It can be decomposed into the intrinsic persistence length $\ell_\text{i}$ and a contribution $\ell_\text{c}$ due to the electrostatic repulsion \cite{chen12, roth18}. Extracting $\ell$ from experiments and simulations typically requires a model and can lead to ambiguous results \cite{hsu13}. Here we extract $\ell$ from the bond correlations \cite{ghobadi16, guerra24}
\begin{equation}
  C(|i-j|) = \langle \hat{u}_i \cdot \hat{u}_j \rangle = e^{-|i-j|\bar r/\ell},
  \label{eq:persisLen}
\end{equation}
where the unit bond vectors $\hat{u}_i$ connect the neighboring beads $i$ and $i+1$ and $\bar r$ is the effective bond length. Fig.~\ref{fig:paraOpti1}(a) shows $C(|i-j|)$ for uncharged ssDNA ($q=0$) as a function of $K_\text{A}$. While the correlations indeed decay exponentially for small distances as predicted from the worm-like chain model \cite{marko95, hays69}, for larger distances the decay slows down. We extract $\ell_\text{i}$ from the initial exponential decay. As expected, the intrinsic persistence length is controlled by $K_\text{A}$, and in Fig.~\ref{fig:paraOpti1}(b) we show that it follows the linear relation $\ell_\text{i}/\bar r\simeq 1.1+0.65K_\text{A}$. Due to the excluded volume, even for $K_\text{A}\to 0$ we thus find a non-vanishing persistence length (roughly equal to the effective bead size $\sigma$). Assuming that the end-to-end distance obeys $\mean{R_E^2}=2\ell\bar r N$ in line with the worm-like chain model (in the limit where $N\gg\ell/\bar r$), we find for $K_\text{A}=0.85$ an effective bond length $\bar r\simeq 1.02\sigma$.

\begin{figure}
  \centering
  \includegraphics{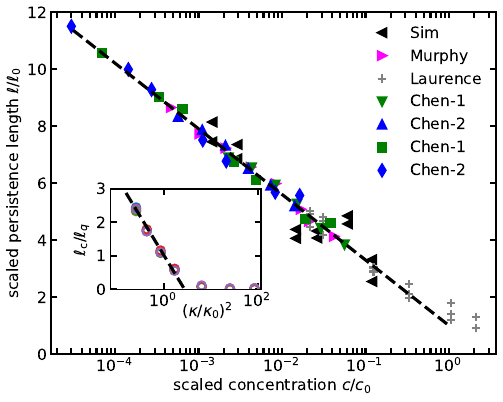}
  \caption{Scaled persistence length as a function of scaled concentrations of counterions for a number of experiments. The dashed line is the scaling function $1-\ln x$. Inset: Simulation results for $\ell_\text{c}$ as a function of $\kappa$ and for several partial charges $q$ in the range from 1 to 2 (colored symbols). The dashed line is again $1-\ln x$.}
  \label{fig:exp}
\end{figure}

The persistence length of ssDNA has been measured in experiments for a range of different sequences and as a function of counterion concentration and valency. While it has been predicted that $\ell\propto I^{-1/2}$ \cite{chen12, odijk77, skolnick77}, in Fig.~\ref{fig:exp} we show that a very good collapse of the available data can be achieved through
\begin{equation}
  \ell(c) = \ell_0f_\ell(c/c_0)
\end{equation}
with scaling function $f_\ell(x)=1-\ln x$. Here, $c$ is the concentration of counterions and $\ell_0$ and $c_0$ are fit parameters so that $\ell(c_0)=\ell_0$. Since $\ell>0$ cannot be negative, this scaling function cannot apply to the limit of large concentrations in which all charges are screened and $\ell\to\ell_\text{i}$ reduces to the intrinsic persistence length. Since for all data sets $\ell>\ell_0$, we conclude that these are within the regime where the persistence length is dominated by electrostatics and that $\ell_0>\ell_\text{i}$ is an upper bound for the intrinsic persistence length. From the values listed in Tab.~\ref{tab:ell}, we see that $\ell_0$ is in the range from 0.2~nm to 1~nm.

\begin{table}[t]
  \begin{tabular}{l|c|r|r}
    \hline
    & Ref. & $\ell_0$ [\AA] & $c_0$ [M] \\
    \hline
    Sim & \citenum{sim12} & 4.0 & 9.6 \\
    Murphy & \citenum{murphy04} & 3.5 & 47.5 \\
    Laurence & \citenum{laurence05} & 11.8 & 0.96 \\
    Chen-1 (Na$^+$) & \citenum{chen12} & 2.8 & 9.4 \\
    Chen-2 (Na$^+$) & \citenum{chen12} & 2.7 & 37.2 \\
    Chen-1 (Mg$^{2+}$) & \citenum{chen12} & 1.9 & 3.9 \\
    Chen-2 (Mg$^{2+}$) & \citenum{chen12} & 2.0 & 9.1 \\
    \hline
  \end{tabular}
  \caption{List of parameters used in the scaling plot in Fig.~\ref{fig:exp}.}
  \label{tab:ell}
\end{table}

In the simulations, the model parameter controlling the range of the electrostatic interactions is $\kappa$ and each bead carries a partial charge $q$. To map between $\kappa$ and concentrations, we employ Eq.~\eqref{eq:kappa} together with the ionic strength $I=\tfrac{1}{2}\sum_ic_iz_i^2$ summing over all ion species with concentration $c_i$, where $z_i$ is the charge number (valency) of the counterions. Typically the concentration $c$ of one species of counterions is changed, for which we obtain $I=I_0+cz^2$ with background ionic strength $I_0$ of the buffer, and the factor of two accounts for the oppositely charged pair of the counterion species in the solution.

We now fix $K_\text{A}=0.85$ and perform simulations varying the partial charge $q$ and extract the persistence length as described. In the inset of Fig.~\ref{fig:exp}, we show that $\ell_\text{c}$ collapses onto
\begin{equation}
  \ell_\text{c} = \ell_qf_\ell((\kappa/\kappa_0)^2)
\end{equation}
with $\kappa_0\simeq0.8$ and prefactor $\ell_q/\bar r\simeq 0.88q-0.47$ that only depends on the partial charge. We have tuned $\ell_q/\bar r$ and $\kappa_0$ so that the increase of $\ell_\text{c}$ for smaller screening $\kappa$ falls onto the same scaling function $f_\ell(x)$. Clearly, the simulations are at the crossover to fully screened charges with $\ell_\text{c}\to0$ and only access a much smaller range of scaled concentrations. Going to smaller $\kappa$ would require to increase the cut-off radius in the simulations, which makes the force calculations computationally expensive. Matching the model with experiments could be achieved through setting $c=c_0(\kappa/\kappa_0)^2$ and matching $q$ (and $\ell_\text{i}$) to obtain the desired $\ell_0$. However, we note that mapping $x=c/c_0=(\kappa/\kappa_0)^2$ together with Eq.~\eqref{eq:kappa} implies
\begin{equation}
  c_0^\text{sim} = \frac{\kappa_0^2}{8\pi\lam_\text{B}N_\text{A}z^2} \simeq \frac{0.06\;\text{M}}{z^2}\frac{\text{nm}^2}{\sigma^2}.
\end{equation}
Assuming $\sigma=1$~nm, this value is two to three orders of magnitude smaller than the experimental values for $c_0$ (cf. Tab.~\ref{tab:ell}). To get closer to the experimental values, we would have to increase $\kappa_0$ through choosing different model parameters. We refrain from doing so and in the remainder, we continue to employ $K_\text{A}=0.85$ together with $r_0=1$, which corresponds to an intrinsic persistence length of $\ell_\text{i}\simeq 1.63$. The partial charges are set to $q=1.5$, smaller than the nominal charge of $2$ per bead (one per nucleotide).

\subsection{Coil-to-globule transition}
\label{sec:coilGlobTran}

\begin{figure}[b!]
  \includegraphics{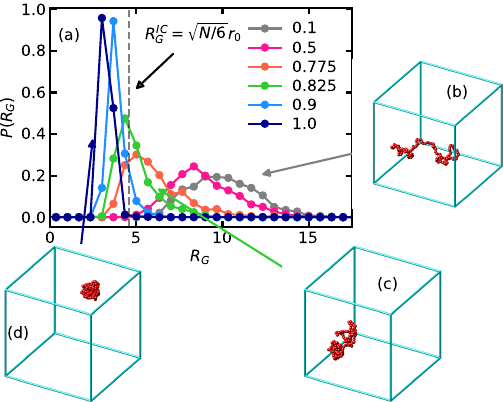}
  \caption{\label{fig:radGyr} Collapse of a single homopolymer. (a) Distribution $P(R_\text{G})$ of the radius of gyration $R_\text{G}$ for $N = 125$ beads at fixed $\kappa = 1.46$ for different values of $\epsilon_\text{H}$ as indicated in the legend. The vertical dashed line denotes the radius of gyration of an ideal polymer of $N = 125$ beads. Decrease in the width of the distribution with increase in $\epsilon_\text{H}$ is a signature of coil-to-globule transition of the ssDNA. Representative snapshots are shown for (b) $\epsilon_\text{H} = 0.1$ (coil phase), (c) $\epsilon_\text{H} = 0.825$ (near the coil-to-globule transition), (d) $\epsilon_\text{H} = 1.0$ (globule phase).} 
\end{figure}

\begin{figure*}[t!]
  \includegraphics{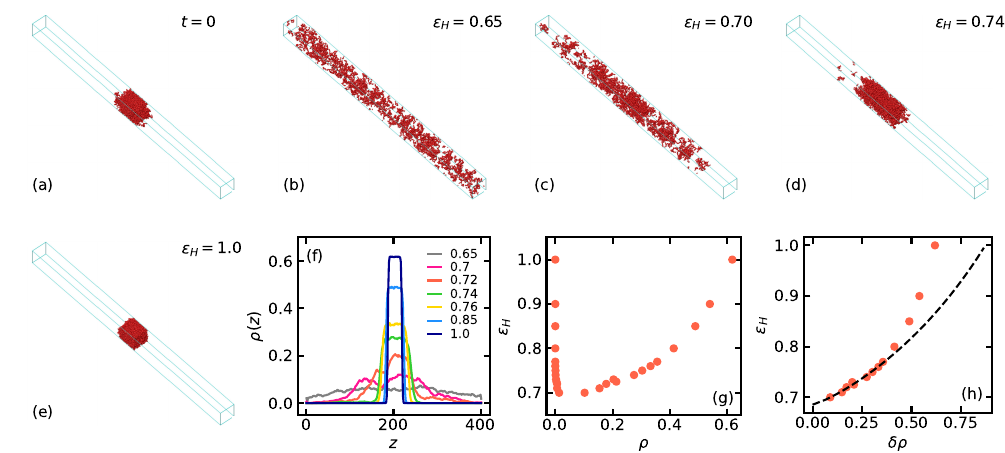}
  \caption{\label{fig:phaseDiagram} Phase coexistence simulation in an elongated box at fixed $\kappa = 1.46$. (a)~Initial configuration ($t = 0$). Final configurations at $t = 250000$ of the system at (b)~$\epsilon_\text{H} = 0.65$ (homogeneous phase), (c)~$\epsilon_\text{H} = 0.7$ (near the phase boundary), (d)~$\epsilon_\text{H} = 0.74$ (in the phase separated region), (e)~$\epsilon_\text{H} = 1.0$ (deep inside the phase separated state, where the low density phase is found to be essentially empty). (f)~Density profile of beads $\rho(z)$ along the elongated axis $z$ for different values of $\epsilon_\text{H}$ (legend). (g)~Coexisting densities (binodal) extracted from the plateau regions of the density profiles $\rho(z)$. (h)~Effective attraction strength $\epsilon_\text{H}$ as a function of the order parameter $\delta \rho = \rho_+ -\rho_-$. The dashed line is the fit to Eq.~\ref{eq:scaling} with the fitted critical attraction strength $\epsilon_\text{H}^\text{c} \simeq 0.69$.} 
\end{figure*}

We now incorporate an effective attraction between beads with free parameter $\epsilon_\text{H}$. We first consider a single polymer and investigate the radius of gyration \cite{baul20}
\begin{equation}
  R_\text{G} = \sqrt{\frac{1}{N}\langle \sum_{i = 1}^{N} (\boldsymbol{r}_i - \boldsymbol{r}_\text{CM})^2 \rangle}
  \label{eq:radGyr}
\end{equation}
to distinguish swollen from a collapsed globule-like state. We follow the same simulation procedure as in Sec.~\ref{sec:params} with the difference of a slightly smaller simulation box $L \approx 40$. In Fig.~\ref{fig:radGyr}(a), we show the probability distribution $P(R_\text{G})$ of the radius of gyration $R_\text{G}$ of a single ssDNA with $N = 125$ beads at fixed $\kappa = 1.46$ for different values of $\epsilon_\text{H}$ together with representative snapshots. At relatively small vales for $\epsilon_\text{H}$, we observe a wide distribution $P(R_\text{G})$ indicating that the ssDNA homopolymer is elongated (swollen). With increase in $\epsilon_\text{H}$ the distribution becomes sharper, indicating that the ssDNA has collapsed. Following established practice \cite{baul20}, we employ the radius of gyration of an ideal (Gaussian) polymer $R_\text{G}^\text{IC} = \sqrt{N/6} r_0$ of equivalent length ($N = 125$ beads) as the threshold value that separates coil-like and globule-like states. The peak of the distribution becomes smaller than $R_\text{G}^\text{IC}$ above $\epsilon_\text{H} \simeq 0.8$.

For fixed $\epsilon_\text{H}$, we determine the probability $P_\text{c}(\epsilon_\text{H})$ that $R_\text{G} > R_\text{G}^\text{IC}$ and $P_\text{g}(\epsilon_\text{H})=1-P_\text{c}$. The coil-like to globule-like transition strength of attraction $\epsilon_\text{H}^{\theta}$ is defined by equating the probabilities of the two states, that is $P_\text{c}(\epsilon_\text{H} = \epsilon_\text{H}^\text{c}) \approx P_\text{g}(\epsilon_\text{H} = \epsilon_\text{H}^\text{c})$, in line with the $\theta$ transition point of conventional thermoresponsive polymers.

\subsection{Direct coexistence simulations}
\label{sec:phaseTran}

We now move beyond a single polymer and consider ensembles of ssDNA homopolymers through direct coexistence simulations of $N_\text{m} = 100$ polymers, each comprising $N = 125$ beads in a strongly elongated simulation box. We fix the box sizes $L_x = L_y = 25$, which are larger than the largest value of the average $\mean{R_\text{G}}$ obtained at the smallest value of $\kappa = 0.46$ considered. The polymers are initialized randomly inside the elongated simulation box. First, we perform an NPT simulation at fixed temperature $T = 1$ and fixed $\epsilon_\text{H} = 0.9$. We apply a pressure $P = 0.025$ along the $z$-axis to determine the average edge length $L_z^0$. Next, we extend the simulation box to $L_z\approx 10 L_z^0$ along the $z$-dimension to obtain the initial configuration ($t = 0$), as shown in Fig.~\ref{fig:phaseDiagram}(a). Then we perform low-friction Langevin dynamics simulations in the fixed (NVT) elongated box for $t = 250000$ and collect the data at uniform interval $\Delta t = 2500$. For $\kappa \leq 0.65$, we consider a larger pressure $P \geq 2.0$ and a larger value of binding free energy $\epsilon_\text{H} \geq 3.0$ to obtain the initial configurations with $L_z \approx 20 L_z^0$.

We set $\kappa = 1.46$ and study the effect of $\epsilon_\text{H}$. At small values of $\epsilon_\text{H}$, the ssDNA polymers start to spread out and exhibit a homogeneous distribution [Fig.~\ref{fig:phaseDiagram}(b) for $\epsilon_\text{H} = 0.65$]. Increasing $\epsilon_\text{H}$, we observe that the homogeneous phase breaks into dense clusters and dilute regions [Fig.~\ref{fig:phaseDiagram}(c) for $\epsilon_\text{H} = 0.7$]. Further increasing the attractive strength $\epsilon_\text{H}$, we find that the clusters merge into a single cluster with only a few ssDNA polymers populating the surrounding low density phase [Fig.~\ref{fig:phaseDiagram}(d) for $\epsilon_\text{H} = 0.74$], before all molecules condense into the dense region [Fig.~\ref{fig:phaseDiagram}(e) for $\epsilon_\text{H} = 1.0$].

In Fig.~\ref{fig:phaseDiagram}(f), we show the time-averaged density profiles $\rho(z)$ of beads over the last $50$ sampled configurations. From these density profiles, we extract the coexisting densities $\rho_\pm$ (the binodal) of the dilute gas and dense cluster from the corresponding plateaus, which are plotted in Fig.~\ref{fig:phaseDiagram}(g). To estimate the critical value of the interaction strength $\epsilon_\text{H}^\text{c}$ at the transition point, we fit the density difference $\delta\rho=\rho_+-\rho_-$ to the power law (in the range $\delta\rho<0.3$)
\begin{equation}
  \delta \rho = B (1 - \epsilon_\text{H}^\text{c} / \epsilon_\text{H})^{\beta}
  \label{eq:scaling}
\end{equation}
with two further fit parameters $\beta$ and $B$. We find that the exponent $\beta$ depends on $\kappa$ and deviates from the Ising value $\beta_0\simeq 0.325$, in particular for small $\kappa$.

\begin{figure}[t]
  \includegraphics{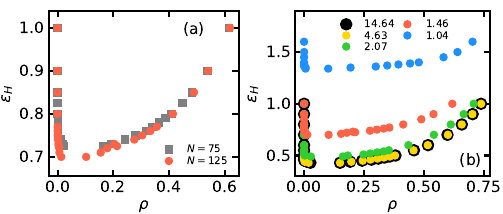}
  \caption{\label{fig:phasedia} Influence of chain length and screening. (a) Comparison of phase diagram for two different size of ssDNA homopolymers, namely $N = 75$ and $N = 125$ beads, as shown in the legend, at fixed $\kappa = 1.46$. The binodal is found to shift towards larger $\epsilon_\text{H}$ with decrease in ssDNA size ($N$ beads). (b) Comparison of the phase diagram for different values of $\kappa$ (legend) for fixed size of ssDNA homopolymers with $N = 125$ beads. Here the binodal moves towards larger $\epsilon_\text{H}$ with decrease in $\kappa$ (or, equivalently, the decrease in the background counterion concentration $c$).}
\end{figure}

In Fig.~\ref{fig:phasedia}(a), we study the effect of chain length. To this end, we compare the phase diagram for two different values, $N = 125$ and $N = 75$, at fixed $\kappa = 1.46$. We consider $N_\text{m} = 166$ homopolymers for $N = 75$, while $N_\text{m} = 100$ polymers for $N = 125$, so that the total number of beads remain close in both cases. With decrease in polymer size, we find that the phase boundary moves towards larger $\epsilon_\text{H}$, which is qualitatively consistent with the increase in lower cloud-point temperature $T^\text{cp}$ of LCST-type polymers, low complexity proteins, and single stranded nucleic acids (DNA and RNA) \cite{kohno15, dignon19, merindol18}.

With decrease in $\kappa$ (equivalently, decreasing the counterion concentration $c$), the electrostatic repulsion between the beads increases. In Fig.~\ref{fig:phasedia}(b), we compare the phase diagram of the ssDNA polymers of fixed size $N = 125$ for different values of $\kappa$, as shown in the legend. With decrease in $\kappa$, our simulation data shows a shift in the phase boundary towards larger $\epsilon_\text{H}$. Hence, larger strength of attraction is required to overcome the effective repulsion between the beads at smaller $\kappa$. The shift in the phase boundary towards larger $\epsilon_\text{H}$ with decrease in $\kappa$ is qualitatively consistent with the experimentally reported increase in $T^\text{cp}$ of ssDNA \cite{merindol18}, RNA \cite{wadsworth23}, and various low complexity proteins \cite{dignon19} with decrease in counterion concentration.

\subsection{Relation to coil-to-globule transition}

Short thermoresponsive polymers of hydrophobic nature are known to exhibit a coil-to-globule transition in the low density limit at the $\theta$-temperature $T^{\theta}$ \cite{dignonPNAS18}. Increasing the size of the polymers is known to reduce $T^{\theta}$. On the other hand, suspensions of polymers with hydrophobic monomers, low complexity proteins with mostly hydrophobic moieties, and single-stranded nucleic acids (both DNA and RNA) are known to exhibit phase separation above a concentration-dependent LCPT ($T^\text{cp}$). Longer polymers are known to favor phase separation by decreasing $T^\text{cp}$ \cite{dignon19, merindol18}.

\begin{figure}[t]
  \includegraphics{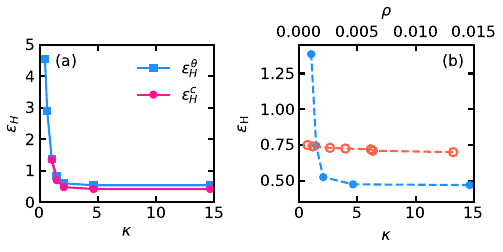}
  \caption{\label{fig:transition} (a)~Critical value $\epsilon_\text{H}^\text{c}$ as a function of $\kappa$ for polymers with $N=125$ beads. We also show the transition $\epsilon_\text{H}^\theta$ to the globular state for a single polymer, which consistently is slightly larger. We do not observe phase coexistence for $\kappa < 1.04$, while the ssDNA homopolymers exhibit a coil-to-globule transition for all values of considered $\kappa$. (b)~Transition value $\epsilon_\text{H}$ as a function of $\kappa$ for fixed global bead density $\rho\simeq7.5\times10^{-4}$ (closed symbols, lower axis) and for fixed $\kappa=1.46$ as a function of $\rho$ (open symbols, upper axis).}
\end{figure}

In Fig.~\ref{fig:transition}(a), we compare the estimated $\theta$ transition strength $\epsilon_\text{H}^{\theta}$ of the coil-to-globule transition of a single ssDNA homopolymer ($N = 125$) with the critical transition strength $\epsilon_\text{H}^\text{c}$ of a collection of ssDNA homopolymers ($N_\text{m} = 100$) of same size ($N = 125$) for different values of $\kappa$. Electrostatic shielding reduces with decrease in $\kappa$, and the long-range electrostatic interaction between the beads increases. Consequently, the ssDNA homopolymers become more swollen at smaller $\kappa$ and the persistence length increases. While a coil-to-globule transition is observed at much smaller $\kappa$, phase coexistence is not observed for $\kappa < 1.04$. Instead, we find small clusters that tend to merge and form connected clusters when further increasing $\epsilon_\text{H}$. For larger $\kappa$ (or, equivalently, larger counterion concentration $c$), $\epsilon_\text{H}^{\theta}$ is found to be larger than $\epsilon_\text{H}^\text{c}$, which is qualitatively consistent with larger $T^{\theta}$ than the critical transition temperature $T^\text{c}$ of similar polymer size. Qualitatively, a larger strength of attraction between the beads is required to counteract the configurational entropy loss of a single polymer in the globule state, while smaller strength of attraction between the beads is required for a multi-polymer system due to the increase in mixing entropy between the polymers because of their entanglement. Interestingly, the difference between $\epsilon_\text{H}^{\theta}$ and $\epsilon_\text{H}^\text{c}$ decreases for smaller $\kappa$.

\subsection{Mapping to experimental phase diagrams}

\begin{figure*}[t]
  \includegraphics{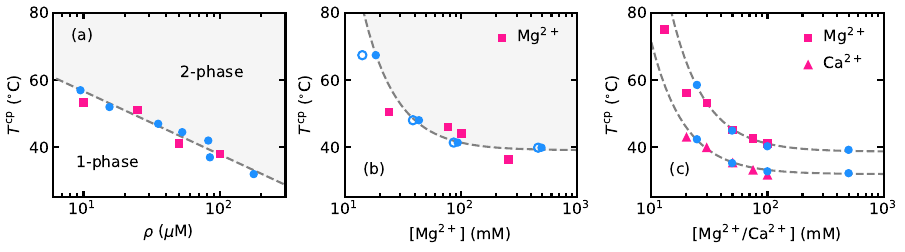}
  \caption{\label{fig:data} Mapping the effective attraction strength $\epsilon_\text{H}$ to experimental cloud points. (a)~Cloud-point temperatures $T^\text{cp}$ ($\blacksquare$) as a function of ssRNA (CAG)$_{31}$ molecular concentration at fixed Mg$^{2+}$ counterion concentration $c = 50$ mM. Data from Wadsworth \emph{et al.} \cite{wadsworth23}. Closed circles ($\bullet$) show the phase boundary of the low density side of the simulated phase diagram for $\kappa = 1.46$, where we map $\epsilon_\text{H}$ to temperature as described in the main text. (b)~Same experiment for fixed molecular concentration $\rho=10$~$\mu$M and varying the concentration of counterions ($\blacksquare$). Circles are from simulations: closed circles indicate the mapping according to Eq.~\eqref{eq:conc} with constant Bjerrum length $\lam_\text{B}$ and open symbols show the shift of concentrations when taking into account the temperature dependence of $\lam_\text{B}$. (c)~Cloud-point temperatures $T^\text{cp}$ of ssDNA poly-adenine plus barcode solutions poly(A$_{20}$-i) at $0.06$ mg ml$^{-1}$ as a function of counterion concentration for two different divalent counterions: Mg$^{2+}$ ($\blacksquare$) and Ca$^{2+}$ ($\blacktriangle$). Data from Merindol \emph{et al.} \cite{merindol18}. Circles are obtained from mapping the same curve $\epsilon_\text{H}(\kappa)$ as in panel (b). In all panels: The dashed gray line is a guide to the eye delineating the homogeneous from the two-phase region.}
\end{figure*}

We have shown that the coarse-grained model qualitatively reproduces phase separation as we increase the model parameter $\epsilon_\text{H}$, which corresponds to an increased gain of binding free energy between beads. Figure~\ref{fig:transition}(b) shows the binodal for the (experimentally relevant) very dilute bead densities. On this scale, the binodal is essentially flat, $\epsilon_\text{H}\simeq\epsilon_\text{H}^\text{c}$. On the contrary, changing $\kappa$ (at constant global $\rho$) leads to a pronounced change of the transition strength $\epsilon_\text{H}$. For small $\kappa$, the repulsive electrostatic interactions become dominant, which requires a strongly growing $\epsilon_\text{H}$ to overcome repulsion between the beads of different polymers to aggregate. At the same time, this is the regime where the persistence length $\ell$ strongly increases.

To make contact with experiments, we now assume that $\epsilon_\text{H}=\epsilon_\text{H}(T)$ is a function of temperature. We make the ansatz $-\epsilon_\text{H}=u-Ts$ in line with a free energy, where $u$ and $s$ are the (roughly time-independent) enthalpic and entropic contributions, respectively. This allows to map $\epsilon_\text{H}$ to a temperature through $T=(u+\epsilon_\text{H})/s$. Experiments typically only access the extreme dilute region $\rho\to0$ of the phase diagram and estimate the binodal from the cloud-point temperature. Figure~\ref{fig:data}(a) shows the cloud points obtained by Wadsworth \emph{et al.} \cite{wadsworth23} for (CAG)$_{31}$ ssRNA at counterion concentration $c=50$ mM. These experiments are performed in a buffer that controls pH and also contains monovalent Na$^+$ counterions at fixed concentration of 25 mM, while the concentration $c$ of divalent Mg$^{2+}$ is varied ($z=2$). Converting the number density of beads $\rho$ to concentration of ssRNA, we see that simulations access the experimentally relevant regime. We map $\epsilon_\text{H}$ to temperature through choosing $s\simeq0.002$ and $u\simeq-0.09$, which agree quite well with the experimentally observed cloud-point temperatures. As anticipated, we require a positive gain $s>0$ of entropy in order to capture the LCST behavior.

In a second data set, the density of ssRNA is fixed to $\rho=10$ $\mu$M while the concentration of counterions is varied, allowing to assess the role of electrostatic interactions. At variance with the model [Fig.~\ref{fig:transition}(b)], the response in experiment (Ref.~\citenum{wadsworth23}) is less strong, with a moderate increase of cloud-point temperature as the concentration is reduced. For a comparision, we now estimate $\epsilon_\text{H}$ along constant $\rho$ for the different values of $\kappa$ (Fig.~\ref{fig:transition}(b)). Consequently, using the previous values for $u$ and $s$ does not result in agreement with the observed temperatures. To map the model parameter $\kappa$ to concentration, we invert Eq.~\eqref{eq:kappa} to obtain
\begin{equation}
  c(\kappa,T) = \frac{\kappa^2}{8\pi\lam_\text{B}(T)N_\text{A}z^2} - \frac{I_0}{z^2}
  \label{eq:conc}
\end{equation}
with $c$ in mM, where here we take into account the background $I_0=25$ mM. We can still attempt to fit the model (taking into account $I_0=25$ mM), but now we obtain the much larger values $u\simeq9.85$ and $s\simeq0.033$. In Fig.~\ref{fig:data}(b), we show the result for fixed $\lam_\text{B}\simeq0.71$ nm and the result when taking into account that $\lam_\text{B}(T)$ in principle also depends on temperature (employing the empirical expression for the dielectric constant of water from Ref.~\cite{malmberg56}). However, the temperature-dependent shift in concentration is smaller than the uncertainties.

In another experiment, Merindol \emph{et al.} \cite{merindol18} have studied phase separation in fluids of ssDNA homopolymers made of adenine as well as block copolymers poly(A$_{20}$-i) composed of an adenine block and a ``barcode'' that can be used for specific binding of the complementary block. Here we focus on the non-specific phase separation, for which data is available that again studies the influence of counterion concentration at fixed molecular weight ($0.06$ mg ml$^{-1}$), which we plot in Fig.~\ref{fig:data}(c). Since the binodal for $\rho$ is essentially flat [Fig.~\ref{fig:transition}(b)], we employ the same curve $\epsilon_\text{H}(\kappa)$ as before and determine $u$ and $s$ to map the temperature that fits the observed behavior. Interestingly, the cloud points depend on the specific ion and Ca$^{2+}$ has lower cloud-point temperatures than Mg$^{2+}$. We can easily capture this ion-specific effect through different enthalpic and entropic contributions to $\epsilon_\text{H}$, with Ca$^{2+}$ implying a much larger entropy $s\simeq 0.09$ ($u\simeq 27.0$) released per bead in comparison with Mg$^{2+}$ having $s\simeq 0.047$ ($u\simeq 14.2$).


\section{Conclusions}
\label{sec:conc}

To conclude, we have studied phase separation in a model of semiflexible charged polymers motivated by recent advances that exploit phase separation in DNA-based fluids. The model combines two nucleotides into one coarse-grained bead, and in the first step we only consider a single bead type. A generic attractive potential between beads models the underlying binding free energy. We thus neglect specific effects from hybridization, although we anticipate that these play an important role in future applications. In future work, we will include nucleobase-specific hybridization through introducing and parametrizing different bead types.

Direct coexistence simulations provide access to the experimentally relevant molecular concentrations. We consider two recent experimental works, using cloud-point temperatures as proxies for the phase transition line. Although the experiments do not strictly employ homopolymers, we capture their phase behavior through a single bead type to obtain first insights into the behavior of model parameters. Moreover, the physical behavior of ssDNA and ssRNA is sufficiently alike on the coarse grain scale of our model.

We find that changing the concentration of counterions has a strong effect on the model phase boundary at variance with the much gentler slope seen across experiments. Mapping temperature to an effective binding free energy $\epsilon_\text{H}$ captures the qualitative trends when changing density and counterion concentration. It is, however, not sufficient to explain the discrepancy, which is resolved through recognizing that counterions not only determine the screening length of the direct electrostatic interactions but can also mediate intermolecular ``bridges'' \cite{wadsworth23}. Moreover, release of counterions constitutes a substantial entropic gain. It is thus not surprising that our effective attraction strength $\epsilon_\text{H}$ is also affected. The next step would be to develop a more comprehensive model for this effect, which would benefit from more experimental data.

Although biomolecular condensates of various proteins and nucleic acids are believed to be liquid-like, nucleobase specific hybridization between the single stranded RNA, present in substantial amount in various biomolecular condensates, could produce a system-wide network \cite{wadsworth23}. Consequently, the liquid-like properties of the biomolecular condensates could be modified into gel-like or solid-like properties that are frequently observed in several neurodegenerative diseases \cite{patel15}. Moreover, nucleic acid hybridization could modify the local densities of the condensates. In our numerical simulations, we observe monotonic increase in the bead density inside the high density phase with increase in attractive potential strength. Additionally, near the boundaries of the high density clusters, we found spatial variation in the bead density. Our work provides a starting point to investigate the diffusive properties of various short and (semi)flexible polymers across various spatial locations of the high density regions \cite{wang23}. We believe that such investigations would provide several valuable insights into the understanding of the behavior of various biomolecular condensates.


\begin{acknowledgments}
  We acknowledge financial support from the Deutsche Forschungsgemeinschaft (DFG) through TRR 146 (grant no. 233630050) and SFB 1551 (grant no. 464588647). SDK would like to thank Swarn Lata Singh for several fruitful discussions. All computations have been performed through bwHPC (HELIX), which is supported by the state of Baden-Württemberg and the Deutsche Forschungsgemeinschaft through grant INST 35/1597-1 FUGG.
\end{acknowledgments}


%

\end{document}